# Computational Complexity of UAP Reverse Engineering
## A Formal Analysis of Automaton Identification and Data Complexity

Karim Daghbouche
GridSAT Stiftung, Hannover (Germany)
karim@gridsat.io

## ABSTRACT

This white paper demonstrates that reverse engineering Unidentified Aerial Phenomena (UAP) is NP-complete under classical computational paradigms. By modeling UAP reconstruction as an automaton identification problem with a *state characterization matrix $M(D, T, E)$* and examining the inherent challenges in data gathering as well as unknown physics, we show that inferring internal mechanisms (such as *Isotopically-Engineered-Materials* or unconventional propulsion systems) from finite observational data is computationally intractable. Data $D$, comprising both operational non-reproducible observations and reproducible analysis data from purported crash retrievals, remains inherently fragmentary. Even if UAP observables were reproducible, the absence of a comprehensive theoretical framework ensures that reverse engineering remains NP-complete, and may escalate to PSPACE-hard or to an *Entscheidungsproblem*. This intractability challenges current UAP reverse engineering efforts and has profound implications for transparency on UAP technology and related venture investments. Hence, UAP are as analogous to modern smartphones in the hands of *Neanderthals.*

---

**CONTENT:**

# 1. INTRODUCTION

## 1.1 Reverse Engineering as a Computational Process

Reverse engineering is the process of deducing the internal design, functionality, and operating principles of a system solely from its observable outputs, in the absence of formal documentation. This process is analogous to "code cracking," where hidden structures (such as encryption keys) are inferred from available data. [1] Reverse engineering typically involves:

   a. *Observation and Data Collection*: Recording input–output pairs under diverse conditions.

   b. *Decomposition and Abstraction*: Identifying system components, interactions, and governing principles.

   c. *Reconstruction and Validation*: Formulating a model that reproduces the observed behavior and testing it against further data.

These techniques are widely used in fields such as integrated circuit (IC) analysis, software decompilation, mechanical system analysis, and in biological research. Notably, even traditional reverse engineering tasks become NP-complete [2] [3] [4] stressing the fundamental computational challenges underlying these endeavors. In contemporary terms, an "automaton" is understood as a state machine model, a concept essential for designing digital circuits, verifying protocols, and modeling complex systems.

## 1.2 Defining Key Terms

*Unidentified Aerial Phenomena (UAP):*

UAP means (A) airborne objects that are not immediately identifiable; (B) transmedium objects or devices; (C) and submerged objects or devices that are not immediately identifiable and that display behavior or performance characteristics suggesting that the objects or devices may be related to the objects or devices described in subparagraph (A) or (B).
The DoD considers Unidentified Anomalous Phenomena (UAP) as sources of anomalous detections in one or more domain (i.e., airborne, seaborne, spaceborne, and/or transmedium) that are not yet attributable to known actors and that demonstrate behaviors that are not readily understood by sensors or observers. [5]

*Non-Human Intelligence (NHI):*

For this white paper, NHI is defined as any non-human entity capable of manifesting with defined UAP - that is, any high-tech system or artifact not produced by human engineering. [6]

*Computational Complexity:*

Computational complexity [7] is often likened to a "zoo" of computational challenges (FIG. 3), encompassing fundamental questions such as P vs. NP [8] and the consequences of their resolution. If P=NP were true [9] [10] [11] [12], it would imply that all problems with solutions verifiable in polynomial time could also be solved in polynomial time. This computational breakthrough would revolutionize technology and lifestyle, enabling feats like cures for all diseases, design automation (hard/software/molecular), (almost) unlimited resources (energy/materials) and formal (mathematical/scientific) knowledge, material design on isotopic level, inter-galactic travel, inertia freedom, weightlessness, non-monetary zero-marginal cost economy [13] [14] etc.), all under the umbrella of its implementation as a Non-Deterministic Processor (NDP). [15]

*Automaton:*

An automaton is an abstract computational model [16] - typically a finite state machine vs. infinite state machines such as *Turing Machines* [17] - that processes sequences of inputs by transitioning among a finite set of states according to a set of predefined rules. Formally, a finite automaton consists of:

a. A finite set of states.

b. An input alphabet, which is a set of symbols.

c. A state transition function that determines how the automaton moves from one state to another based on the current state and an input symbol.

d. An initial state, where the processing begins.

e. (Optionally) A set of accepting states or an output function, depending on the application.

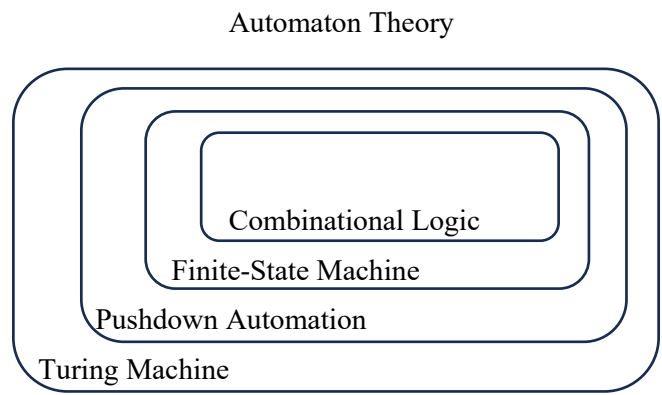

FIG. 1   Classes of Automata

Automata serve as an effective model for reverse engineering:

*A) Simplification of System Behavior*: Engineered systems, such as digital circuits, communication protocols, and embedded controllers, exhibit behavior that can be abstracted as a series of discrete state transitions. A finite automaton captures these transitions succinctly, making it a natural framework for modeling the operational behavior of a system.

*B) Modeling Input–Output Relationships*: Since reverse engineering aims to deduce the internal structure of a system from its observable outputs in response to given inputs, automata are designed to represent these input–output relationships, allowing engineers to formalize and analyze the system's behavior systematically.

*C) Finite Complexity*: Although the real-world systems might be complex, many of the reverse engineering challenges involve finite and discrete behaviors. Finite automata provide a tractable model for these behaviors, allowing the application of computational complexity theory.

*D) Theoretical and Practical Insights*: Using automata as the underlying model allows for a rigorous analysis of reverse engineering challenges. The formal results provide insights into the computational limits of reverse engineering tasks and guide the development of heuristic and approximate methods when exact solutions are computationally prohibitive.

In summary, automata offer a simplified, yet powerful abstraction of system behavior that directly aligns with the goals of deducing internal structures from observable data. This model not only facilitates a theoretical analysis of computational complexity but also has practical relevance in the design, verification, and analysis of modern engineered systems. [18]

*Observables (Data D):*

The data utilized in UAP reverse engineering can be classified into two categories:

*A) Operational Non-Reproducible Observations*: These are field observations that are often captured passively by sensors and include: [19]

- *Positive Lift without Flight Surfaces*: Observations indicating lift generation without conventional aerodynamic structures.
- *Sudden/Instantaneous Acceleration*: Reports of rapid acceleration beyond known mechanical limits.
- *Hypersonic Velocity Without Signatures*: High-speed travel without the typical byproducts, such as sonic booms or heat trails.
- *Trans-Medium Travel*: The ability to transition seamlessly between different environments (e.g., air, water, space).
- *Low Observability or Cloaking*: Cases where objects are extremely difficult to detect through standard sensory or sensor means.
- *Biological Effects on Humans and Animals*: In some instances, close encounters have been associated with unexplained physiological or psychological impacts.

*B) Reproducible Analysis Data*: This category comprises data gathered from purported crash retrievals [20] [21] or dedicated laboratory analyses. Such data is structured and includes detailed material characteristics, such as:

- *Isotopically-Engineered-Materials*: Engineered materials that exhibit properties not found in naturally occurring substances. These may be designed at the isotopic level, i.e., the material's performance is finely tuned through its isotopic composition[*], yielding unique electromagnetic, acoustic, or mechanical properties. [22]
- *Systemic Measurements*: Data related to classic mechanics such as weight, dimensions, energetic calculations for acceleration and deceleration, the internal configuration, structural integrity, and material composition derived from recovered UAP artifacts.

*Test States (T) and Experiments (E):*

In the automaton identification framework [23], $T$ represents the set of test inputs (assumed to be prefix-complete) and $E$ represents the set of experiments (assumed to be suffix-complete) used to construct the *state characterization matrix*.

*Prefix-completeness:*
A set of strings is said to be prefix-complete if for every string in the set, every prefix of that string is also included in the set. In the context of automaton identification, if the set $T$ of test states is prefix-complete, then for any input string in $T$, all of its leading segments (or prefixes) are also present in $T$. This ensures that intermediate steps of state transitions are captured, which is crucial for accurately reconstructing the system's behavior.

*Suffix-complete:*
Similarly, a set of strings is suffix-complete if for every string in the set, every suffix of that string is also contained in the set. When the set $E$ of experiments is suffix-complete, it means that for any experimental input used to probe the system, all of its trailing segments (or suffixes) are included. This property guarantees that the outputs corresponding to the latter portions of input sequences are observed, thereby filling in critical details needed to fully characterize the system.

Together, these properties ensure that the *state characterization matrix* is as complete as possible, capturing all partial sequences that occur during the system's operation.

---

[*] With about 80 natural elements, engineering isotopes exhibits approximately 253 distinct atoms where the additional degree of freedom introduced by isotopic variation exponentially multiplies the number of possible configurations, hence the design space analogous to the transitions between the *Stone Age*, to the *Bronze Age*, and eventually to the current *Iron Age*, where each leap was driven by a qualitatively new level of control over materials. Even a slight variation in isotopic composition can lead to radically different physical properties.

*Example:* We imagine a simple music box that produces a sequence of musical notes as it rotates. The music box's mechanism can be modeled as an automaton - a finite state machine that transitions between a finite set of states based on its internal configuration. Reverse engineering the music box means deducing its internal state transitions (which pins trigger which notes) from observed outputs (the notes played).

Suppose we want to reconstruct the music box's mechanism using a *state characterization matrix*. We define:

*Test States (T)*: Let $T$ be the set consisting of state $A$ and state $B$, where "$A$" and "$B$" represent two distinct positions or conditions of the music box's cylinder.

*Experiments (E)*: Let $E$ be the set consisting of experiment $X$, experiment $Y$, and experiment $Z$, where these experiments represent three successive positions on the cylinder at which the comb interacts with the pins.

*Observational Data (D)*: For each combination of a test state (or a state derived from a test state) and an experiment, we record the corresponding output. In this music box example, the output could be represented by the note played (or a binary indicator representing whether a particular note is produced).

Now, we construct the *state characterization matrix $M(D, T, E)$*. Each row corresponds to a test state (or a state reachable from $T$), and each column corresponds to an experiment. E.g., suppose the matrix $M$ is as follows:

```
          X    Y    Z
         -----------------
Row A:    1    0    1
Row AA:   1    0    1      ← This row is "tied" to Row A (identical outputs)
Row B:    0    1    0
```

FIG. 2　State Characterization Matrix

*Prefix-Completeness T*:　The set $T$ is assumed to be prefix-complete, meaning that if a state (like "$AA$") is derived from a test state "$A$", then the necessary prefixes (in this case "$A$") are already present in $T$.

*Suffix-Completeness E*:　The experiments $E$ are assumed to be suffix-complete, ensuring that for any experimental input, all its relevant suffixes are also included. Here, since our experiments are simply labeled $X$, $Y$, and $Z$, the completeness condition ensures that all parts of the experiment sequence are recorded.

*Tied Rows*:　Notice that the row for "$AA$" has exactly the same output values (1, 0, 1) as the row for "$A$." According to the *Data Matrix Agreement Theorem* in the proof summary, tied rows must have identical entries. This ensures consistency in the model reconstruction: If two states are meant to represent the same internal behavior, their observed outputs must match.

Accordingly, if $M(D, T, E)$ contains no "holes" (i.e., every required observation is present), and if $T$ is prefix-complete and $E$ is suffix-complete, then the reconstructed automaton will agree with all observed data from the music box.

This simple reverse engineering oft a music box illustrates how an automaton can model a system's behavior through a complete and structured *state characterization matrix*. Such a model is the basis for understanding more complex reverse engineering problems.

## 2. AUTOMATON IDENTIFICATION: SUMMARY AND PROOF LOGIC

*E Mark Gold* demonstrated in 1978 [23], that reconstructing a finite automaton from a finite set of observations is NP-complete. The proof is built upon the following key theorems:

### 2.1 Theorem 1: Data Matrix Agreement

A *state characterization matrix* (*M* of *D*, *T*, and *E*) yields a valid automaton if and only if the test states *T* are prefix-complete and the experiments *E* are suffix-complete. Under ideal conditions with complete data, this guarantees that the reconstructed model accurately reflects the system's behavior.

### 2.2 Theorem 2: Transition Assignment is NP-Complete

Determining whether there exists a finite automaton with a specified number of states that agrees with *D* is NP-complete.

Proof Outline:

1. An arbitrary Boolean formula in conjunctive normal form [24] is reduced to the transition assignment problem by mapping each Boolean variable to a state and each clause to an observation in *D*.

2. Verification that a candidate automaton meets these constraints is possible in polynomial time.

3. Since SAT [24] is NP-complete, the reduction establishes that the transition assignment problem is NP-complete.

### 2.3 Theorem 3: Minimal Set of Test States May Not Be Minimum

Even if a feasible set *T* of test states exists that allows for the reconstruction of the automaton, it may not be the smallest possible set. In other words, heuristic approaches may yield a test state set that is sufficient but not optimally minimal, further complicating the problem.

### 2.4 Theorem 4: Automaton Identification-in-the-Limit with Polynomial Time and Data

Under the condition that sufficient and well-structured data is available, a timid state characterization algorithm can eventually identify the automaton in polynomial time, though it might not achieve a minimal state realization. In practice, however, the data requirements grow exponentially, making the task intractable.

## 3. UAP REVERSE ENGINEERING

The process of UAP reverse engineering in automaton identification observes all three stages:

*A) Observational Data (D):*

Comprised of the key UAP observables - positive lift without flight surfaces, sudden acceleration, hypersonic velocity without signatures, trans-medium travel, low observability or cloaking, and biological effects on humans and animals.

*B) Hidden States (S):*

Represent the unknown internal configurations of UAP systems, such as specific isotopic material configurations or novel propulsion mechanisms.

*C) Test States (T) and Experiments (E):*

Represent the controlled inputs and experimental conditions under which UAP behavior is observed. In practice, both observational and crash retrieval data are incomplete, leading to gaps in the *state characterization matrix*.

Direct Mapping:

*Data Matrix Agreement (Theorem 1):*

In UAP reverse engineering, constructing a complete *state characterization matrix* is hindered by incomplete data, even when crash retrievals provide detailed material information.

*Transition Assignment (Theorem 2):*

The process of deducing transitions between hidden UAP states from the observable constraints is equivalent to solving a SAT instance, establishing NP-completeness.

*Minimal Test States (Theorem 3):*

Selecting an optimal set of experimental conditions is computationally challenging and may result in a non-minimal test state set, thereby enlarging the search space.

*Identification-in-the-Limit (Theorem 4):*

Notwithstanding structured data from crash retrievals, the exponential data requirements driven by unknown physics may escalate the problem beyond NP-completeness.

## 4. ESCALATION BEYOND NP-COMPLETENESS

The UAP reverse engineering task intractability escalates mainly because of:

### 4.1 Unknown Physics

UAP systems may operate on exotic or unknown physical principles, resulting in an unbounded or infinite state space. This lack of theoretical constraint can elevate the problem into PSPACE-hard or undecidable domains, where no polynomial-time algorithm exists.

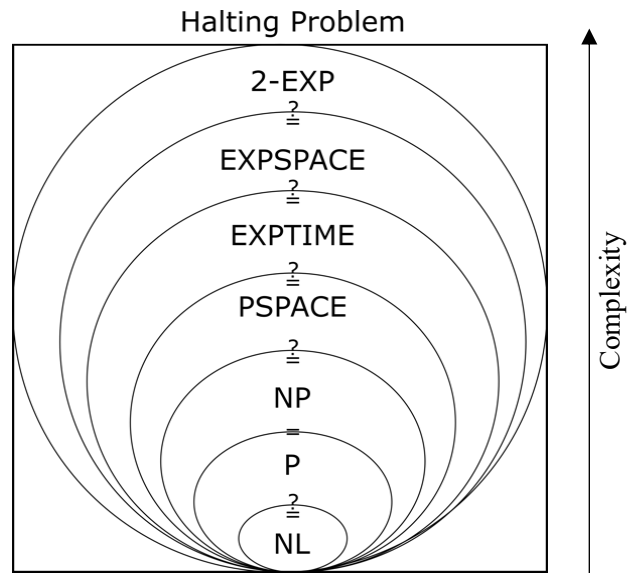

FIG. 3     Complexity Zoo

### 4.2 Severely Fragmentary Data

Incomplete data forces the need for exhaustive, exponential search to "fill in the gaps" of the *state characterization matrix*, further intensifying the computational challenge and potentially pushing the problem into higher complexity classes.

### 4.3 Meta-Computer Architectures

Notwithstanding of UAP data completeness and reproducibility, the underlying computational challenge of reverse engineering UAP systems may escalate far beyond NP-completeness due to the following speculative factors:

*1. Isotopically-Engineered Materials:*

Contemporary material science has yet to advance to the point where atomic bonding can be controlled to produce macroscopic objects with tailored properties. As for UAP we hypothesize that such materials are engineered at the isotopic level to achieve extraordinary mechanical, electromagnetic, and quantum properties. The precise control over atomic composition and bonding poses an immense contemporary challenge, as it requires managing an exponentially vast configuration space, which directly contributes to the intractability of reverse engineering these materials.

*2. Beyond von Neumann Architectures:*

We further assume that these isotopically-engineered materials incorporate an inherent circuitry that functions beyond conventional von Neumann computer architectures [25]. Instead of relying on separate memory and processing units, the material itself could be designed to perform computations intrinsically, integrating signal processing and decision-making at the atomic level. This "meta von Neumann computer architecture" is hypothesized to embody functionalities akin to fundamental data structures [15], where every computable function is pre-materialized and can be instantly accessed via the quantum states of the material.

*3. Comparative Complexity:*

Current concepts of quantum computation are relatively primitive compared to what might be achieved by such advanced materials. If these meta-computer architectures exist, they would operate with a level of efficiency and parallelism far exceeding that of our present-day high-performance computing systems. In effect, our conventional computational models (including quantum computers which are constrained by NP-complete- and hardness themselves [26] [27]) may appear as rudimentary as medieval automata when compared to the potential capabilities embedded in these sophisticated UAP systems.

*4. Assumption on NHI Capabilities:*

We finally assume NHI mastering NDP [15] as a necessary condition for manifesting with UAP [6]. NDPs efficiently manage tasks that are computationally intractable by classical methods.

### 4.4 Increased Complexity: PSPACE-hardness and Undecidability

*PSPACE-hardness*: When the state space is exponentially large or unbounded, the search for a model that agrees with the observed data $D$ (including the "hole filling" in the *state characterization matrix*) becomes a problem that requires exploring a search space that grows with the length of the input. Such problems are often PSPACE-hard, meaning they require an amount of memory (or space) that is polynomial in the input size but can take exponential time.

*Undecidability*: In the extreme case, if the system behaves like a *Turing Machine* [17] (owing to its unbounded state space and integrated meta-computational capabilities), then the reverse engineering task includes *Entscheidungsprobleme* [28] equivalent to the *Halting Problem* [29], which is undecidable. In such cases, no algorithm can determine, for every possible UAP, whether a correct reconstruction exists.

Eventually, the escalation from NP-completeness to PSPACE-hard or undecidable arises from the transition of the model:

*Finite Automata*: NP-complete under Gold's framework.

*Unbounded/Infinite-State Systems*: Reverse engineering such systems inherits the full complexity of *Turing machine* decision problems leading to PSPACE-hardness or yielding an *Entscheidungsproblem*.

Thus, for UAP reverse engineering, where the underlying physics may allow for an unbounded state space (via, e.g., isotopically-engineered materials with integrated meta-circuitry), the computational problem exceeds NP-completeness. The additional complexity required to "fill in the gaps" of fragmentary data in such a system forces the problem into higher complexity classes.

## 5. DATA GATHERING SCENARIOS

Data quality and completeness are critical to any reverse engineering effort. In the context of UAP reverse engineering, data is collected from two primary sources:

### 5.1 Observational Data

- *Nature*: Data is gathered passively from sensor systems (e.g., radar, optical, infrared) across diverse environments.
- *Characteristics*: Fragmentary sensor readings often intermittent due to environmental and technical limitations.
- *Incomplete*: The absence of controlled, systematic experiments results in numerous gaps (holes) in the *state characterization matrix*.

Implication: Even if UAPs exhibit their key operational observables, the resulting dataset is inherently incomplete, limiting our ability to reconstruct the underlying mechanisms.

### 5.2 Crash Retrievals

- *Nature*: Physical artifacts recovered from purported UAP incidents, subjected to detailed laboratory analysis.
- *Characteristics*: Purported crash retrievals yield comprehensive material and structural information, including the properties of materials.
- *Contextual Limitations*: Such purported artifacts imply the lack dynamic operational data (e.g., real-time propulsion behavior) and may additionally be compromised by damage or contamination.

Implication: Despite providing richer datasets, crash retrievals do not capture the full operational behavior of UAP systems, thereby maintaining the intractability of the reverse engineering problem.

### 5.3 Hypothetical Reproducibility

- *Scenario*: NHI perform a flight show in which UAP observables such as positive lift without flight surfaces, sudden acceleration, hypersonic velocity without signatures, trans-medium travel, low observability, and even biological effects are consistently and reproducibly exhibited. In this scenario, while the observables are reliably reproduced in public demonstrations, this reproducibility pertains solely to the external, measurable phenomena. It does not equate to a controlled experimental environment where every variable is isolated and manipulated; rather, it simply confirms that these UAP characteristics occur under the same conditions.

Implication: Although such reproducibility reinforces the observable data, it does not reveal the underlying physical or technological principles. The absence of a comprehensive theoretical framework ensures that the reverse engineering problem remains NP-complete, or may escalate to PSPACE-hard or yield an *Entscheidungsproblem*.

The critical point is that reproducible observables, even when clearly documented during an NHI flight show, do not reveal the underlying physical or technological principles governing UAP systems. Without a comprehensive theoretical framework that explains these principles such as advanced models for isotopically-engineered meta-materials or non-classical propulsion mechanisms, the reverse engineering problem remains fundamentally intractable. Thus, reproducibility in observables reinforces the reliability of the external data, but it does not reduce the inherent computational complexity of reconstructing the internal system.

## 6. CONCLUSION

The automaton identification framework rigorously demonstrates that reconstructing a finite automaton from finite observations is NP-complete. When this framework is applied to UAP reverse engineering, the following must be observed:

*NP-Completeness:*

The reconstruction of UAP systems from a finite set of observables is NP-complete due to the inherent complexity of transition assignment.

*Escalation of Complexity:*

The presence of unknown physics and fragmentary data may escalate the problem into PSPACE-hard or *Entscheidungsprobleme*.

*Data Gathering Challenges:*

Whether data is gathered through passive observations or crash retrievals (which provide detailed material information), the inherent incompleteness of the data prevents a full reconstruction of UAP systems.

*Reproducibility is Insufficient:*

Even if UAPs reproducibly exhibit their operational observables, this reproducibility only reinforces the external constraints without revealing the underlying physical or technological principles. The reverse engineering problem remains computationally intractable.

*Implications for Technology Classification and Investment:*

Given the NP-completeness and potential escalation beyond, efforts to maintain UAP technology as classified for technological supremacy are fundamentally absurd. Likewise, UAP-tech venture funds targeting Frontier Tech would face an insurmountable reverse engineering challenge with contemporary methods.

UAP reverse engineering is a computationally intractable problem within current theoretical frameworks. Notwithstanding structured, reproducible data from crash retrievals or controlled experiments, the combination of NP-completeness, unknown physics, and incomplete data ensures that classical reverse engineering methods cannot efficiently reconstruct UAP systems. They are as analogous to modern smartphones in the hands of *Neanderthals*. While the exotic origin may be recognizable, the underlying functionality, material design, and internal logic with integrated circuits, software protocols, and sophisticated control algorithms are completely alien. Moreover, the supporting infrastructure that underpins modern technology (such as high-speed internet connectivity, fiber optics, satellite networks like Starlink or 5G/WiFi systems) would entirely be beyond *Stone Age* understanding.

## 7. FINAL NOTE

The computational intractability of UAP reverse engineering cannot be overcome by conventional heuristics such as "pencil-and-paper" techniques. Only NDP techniques have the potential to reduce NP-complete problems to polynomial-time complexity of possible state transitions.

Most important, NDP provides the critical framework to "fill in the holes" in the *state characterization matrix* by integrating new mathematical and physical frameworks to systematically address the gaps and uncertainties inherent in fragmentary observational data. While classical reverse engineering efforts remain fundamentally doomed by their inability to manage exponential complexity and incomplete data, only a transformative approach based on NDP can yield efficient and reliable reconstructions of UAP systems.

## 8. ACKNOWLEDGEMENT

This discussion is orchestrated and maintained by *GridSAT Stiftung* - Germany - [gridsat.io](gridsat.io) - gridsat.eth/ and supported by *3onic Systems, Inc.* - Germany - [3onic.com](3onic.com)